\documentclass[a4paper,12pt,final]{article}
\usepackage[english]{babel}
\usepackage[utf8]{inputenc}
\usepackage{t1enc}
\usepackage{floatflt}
\usepackage{graphicx}
\usepackage{psfrag}
\usepackage{bbm}
\usepackage{amsmath}
\usepackage{amssymb}
\usepackage{showkeys}
\usepackage{hyperref}
\usepackage{ifthen}
\usepackage{subfigure}
\usepackage{epstopdf}

\title{Stabilisation of semilocal strings by dark scalar condensates}
\author{Péter Forgács\textsuperscript{1,2}, Árpád Lukács\textsuperscript{1}\\
{\small {}\textsuperscript{1} Wigner RCP RMI, H1525 Budapest, POB 49}\\
{\small {}\textsuperscript{2} LMPT CNRS UMR7350, Universit\'e de Tours, Parc de Grandmont, 37200 Tours, France}
}

\def\d{\mathrm{d}}
\def\e{\mathrm{e}}

\def\kihagy#1{}

\newcommand{\arxiv}[2][]{%
  \ifthenelse{\equal{#1}{}}{%
    \href{http://arxiv.org/abs/#2}{\texttt{arXiv:#2}}%
  }{%
    \href{http://arxiv.org/abs/#2}{\texttt{arXiv:#2 [#1]}}%
  }%
}%

\newcommand{\be}{\begin{equation}}
\newcommand{\ee}{\end{equation}}

\begin{document}
\maketitle

\begin{abstract}
Semilocal and electroweak strings are well-known to be unstable against unwinding by the condensation of the second Higgs component in their cores. A large class of current models of dark matter contains dark scalar fields coupled to the Higgs sector of the Standard Model (Higgs portal) and/or dark U(1) gauge fields. It is shown, that Higgs-portal-type couplings and a gauge kinetic mixing term of the dark U(1) gauge field have a significant stabilising effect on semilocal strings in the ``visible'' sector. 
\end{abstract}

Cosmic strings and their observational signatures have been studied since a long time as they are expected to form in the early universe \cite{kibbleorig, VS, kibble, VachaspatiSP, Ringeval}. 
Even if by now it seems unlikely that cosmic strings could have significantly contributed to structure formation in the universe, string-like excitations in the Standard Model (SM) continue to be of great interest not only from a theoretical point of view, but such objects may eventually leave observable signatures, e.g., in the Large Hadron Collider \cite{semilocal, Nambu, Huang}.
Remarkable string solutions have been uncovered in the bosonic sector of the Glashow-Salam-Weinberg (GSW) theory (In this paper we shall refer to a generalisation of the electroweak sector of the SM allowing its parameters to take on non-physical values as the GSW theory.), for a review see Ref.\ \cite{semilocal}.
A rather interesting class of models emerges by taking the $\theta_{\rm W} \to \pi/2$ limit of the GSW theory, where $\theta_{\rm W}$ denotes the electroweak mixing angle. One obtains this way an Abelian Higgs model with an extended scalar sector having an $SU(2)_{\text{global}}$
symmetry acting on the Higgs doublet, this a a prototype of semilocal models.
Its strings solutions are referred to as semilocal strings  \cite{semilocal, vac-ach, hin1, hin2} and these are quite instructive to study
as being potentially important object in the GSW theory. An important criterion for the physical relevance of string-type objects is their classical stability.  
Semilocal strings turned out to be stable only when the mass of the scalar particle is smaller than that of the (single) gauge boson, as shown in Refs.\ \cite{hin1, hin2}. The stability of electroweak strings (whose progenitors are the semilocal ones) has been considered in Refs.\ \cite{semilocal, JPV1, JPV2, Perkins, GHelectroweak}; it was found that for physically realistic values of $\theta_{\rm W}$, electroweak strings are unstable. 

Moreover there are good reasons to consider extended versions of the GSW theory by a dark sector (DS), motivated by the mystery of dark matter.
In such extended models the question of the possible r\^ole of strings appears naturally. A minimalistic extension of the GSW theory is to couple a (dark) scalar field to the by now firmly established Higgs sector of the GSW theory (Higgs portal) \cite{SilveiraZee, PW}, but there are also well motivated extensions of the GSW theory containing U(1) gauge fields in the DS \cite{ArkaniHamed, ArkaniHamed2}. 
In Refs.\ \cite{VachaspatiDS, Vachaspati1, Vachaspati2, Vachaspati3, Holdom2, HindmarshDS} physical properties and possible observational signatures of cosmic strings in the DS (dark strings) have been considered. A more detailed investigation  of string solutions in Abelian Higgs theories modelling a ``visible'' and a ``dark'' U(1) gauge sector was presented in Ref.\ \cite{HartmannArbabzadah}.  In subsequent works \cite{BrihayeHartmann,BabeanuHartmann} semilocal-type  models with a ``visible'' and a ``dark'' U(1) gauge field  spontaneously broken in both sectors have been investigated. It has been observed that the stability region of semilocal string solutions with a non-zero winding number in the DS can be extended in function of the couplings between the visible and the DS. Higher winding vortices in the U(1)$\times$U(1) model and its supersymmetric generalisation have been considered in Refs.\ \cite{Fidel1, Fidel2}. An earlier work on string solutions in a portal-type theory is Ref.\ \cite{Peter2}.
In all these works only strings with non-zero winding in the DS have been considered, because of the known instabilities of ``visible'' semilocal strings.

The main goal of the present paper is to
complement these studies on dark strings by concentrating on the influence of the DS on ``visible'' semilocal-type string solutions (i.e.\ with zero winding in the DS). We consider a U(1)$\times$U(1) Abelian Higgs (AH) model, whose scalar sector consist of a complex Higgs doublet with (global) SU(2) symmetry coupled to a dark scalar field with an U(1)$\times$U(1) symmetric potential, which is a simple generalisation of the model of Witten \cite{Witten}.
We use this simplified  model to study the effect of the DS on semilocal strings.
It is convenient to distinguish between two symmetry breaking patterns; either both the visible Higgs and the dark scalar field have non-zero vacuum expectation values (2VEV), or there is no symmetry breaking in the DS (1VEV).
The case with 1VEV is directly relevant to the Higgs portal (scalar phantom) model of Refs.\ \cite{SilveiraZee, PW},
whereas when the DS contains gauge fields to model interaction among the dark matter particles the symmetry must be broken in both the visible
and the dark sector (2VEV case) \cite{ArkaniHamed, ArkaniHamed2, VachaspatiDS}.

Generically, semilocal strings are unstable with respect to condensation of the dark scalar field at their core [we shall refer to such strings as dark core, (DC) ones]. In the absence of the gauge kinetic mixing, the DC strings investigated in the present paper, correspond to embeddings of the solutions previously 
found in Refs.\ \cite{Peter} resp.\ \cite{FLCC, FLCC2} into the SU(2)$\times$U(1) symmetric semilocal model coupled to a DS. When the gauge kinetic mixing is different from zero the string solutions we consider here differ from those of Ref.\ \cite{BrihayeHartmann,BabeanuHartmann} in that our strings
have nontrivial winding only in the visible sector.
Our main result is the stability of DC strings with respect to small perturbations for a rather large parameter domain.

It has to be pointed out that a number of mechanisms to stabilise semilocal strings have already been investigated.
In Ref.\ \cite{VachaspatiWatkins}, a stabilising effect due to a bound state of an additional scalar field on semilocal and electroweak strings
has been found. In Ref.\ \cite{PerivolaropoulosPlatis}, it has been shown that a special (dilatonic-type) coupling between the gauge and the scalar field also has a stabilising effect on semilocal strings.

In the complementary limit of the electroweak theory, $\theta_{\rm W} \to 0$, it has been demonstrated that quantum fluctuations of a heavy fermion doublet coupled to the string can also lead to stabilisation in Refs.\ \cite{Weigel1, Weigel2}. Stabilisation of electroweak strings due to the interaction with thermal photons has been demonstrated in Ref.\ \cite{nagabrand}.

The plan of the paper is as follows: in Sec.\ \ref{sec:model} we introduce the models considered, which is followed by the discussion of visible straight string solutions in the 2VEV case and their stability properties in Sec.\ \ref{sec:2VEV}. Next we analyse the 1VEV case in Sec.\ \ref{sec:1VEV}. We conclude in Sec.\ \ref{sec:conclusions}. Some details have been relegated to various appendices: scalar masses in the 2VEV case to Appendix \ref{app:masses}, radial equations of vortices to Appendix \ref{app:radeq} and the linearisation of the field equations about the vortices to Appendix \ref{app:linear}.

\section{Simple models of dark matter}\label{sec:model}
In Refs.\ \cite{ArkaniHamed, ArkaniHamed2}, a unified model of dark matter has been presented, which posits a DS with a U(1) gauge symmetry, spontaneously broken in order to avoid long range interactions. The DS is modelled by an AH model $(C_\mu,\chi)$, where the dark scalar field, 
$\chi$, couples to the GSW theory through a Higgs portal coupling \cite{SilveiraZee, PW}
and the dark gauge field $C_\mu$ through a gauge kinetic mixing term \cite{Holdom}.

We consider the following semilocal model coupled to a DS defined by the Lagrangian\footnote{We use the metric signature $+---$.}:
\begin{equation}
 \label{eq:LagPhi}
 \mathcal{L} = -\frac{1}{4}F_{\mu\nu}F^{\mu\nu}-\frac{1}{4}H_{\mu\nu}H^{\mu\nu}+\frac{\epsilon}{2}H_{\mu\nu}F^{\mu\nu}
 + D_\mu \Phi^\dagger D^\mu\Phi + ({\tilde{D}}_\mu \chi)^*{\tilde{D}}^\mu \chi- V(\Phi,\chi)\,,
\end{equation}
where $\Phi=(\phi_1, \phi_2)$,  $D_\mu \Phi = (\partial_\mu -i A_\mu)\Phi$, ${\tilde D}_\mu \chi = (\partial_\mu -iqC_\mu)\chi$, $F_{\mu\nu}=\partial_\mu A_\nu-\partial_\nu A_\mu$,
$H_{\mu\nu}=\partial_\mu C_\nu - \partial_\nu C_\mu$.
The potential, $V(\Phi,\chi)$ is a slight generalisation of that of the Witten model \cite{Witten},
\begin{equation}\label{eq:pot}
  V(\Phi,\chi) = \frac{\beta_1}{2} (|\Phi|^2-1)^2 + \frac{\beta_2}{2} |\chi|^4 + \beta' |\Phi|^2|\chi|^2 - \alpha |\chi|^2\,.
\end{equation}
The parameters $\beta_1$, $\beta_2$, $\beta'$, $\alpha$ are restricted by demanding
that $V(\Phi,\chi)>0$ for $|\Phi|^2\,,|\chi|^2\to\infty$, resulting in: $\beta_1>0$, $\beta_2>0$, and $\beta'>-\sqrt{\beta_1\beta_2}$.
For a description of the vacua of $V(\Phi,\chi)$ we refer to Refs.\ \cite{FLCC, FLCC2}. The parameters, $\beta'$ and $\epsilon$, correspond to the Higgs portal and gauge kinetic mixing \cite{Holdom}, respectively.

The above model \eqref{eq:LagPhi} can be viewed as the $\theta_{\rm W}\to\pi/2$ limit of the GSW theory coupled to a DS, therefore we shall refer to the fields $\Phi$ and $A_\mu$ as the ``visible sector''; and to $\chi$ and $C_\mu$ as the DS. Apart from the local U(1)$\times$U(1) it has a global SU(2) symmetry acting on the (complex) Higgs doublet, $\Phi$, and we shall refer to \eqref{eq:LagPhi} to as ``semilocal-DS'' model.

In the 2VEV case, for $\epsilon=0$, the gauge boson masses are given as $m_A^2=2\eta_1^2$ and $m_C^2=2q^2 \eta_2^2$,
 where the VEVs $\eta_1$ and $\eta_2$ expressed in terms of the parameters of the potential are listed in Appendix A, Eq.\ \eqref{eq:2VEV}. The scalar particles $\phi_1$ and $\chi$ mix, the analysis thereof is presented in Appendix \ref{app:masses}. The field $\phi_2$ remains massless (in the GSW theory, it is the would-be Goldstone boson corresponding to the longitudinal component of $W^\pm$). For a detailed analysis of the effects of the gauge kinetic mixing we refer to Refs.\ \cite{Vachaspati1, Holdom}.
Unless the mass of the DS scalar $\chi$ is large ($m_\chi \gg 1$ TeV) compared to SM masses, $\epsilon \lesssim 10^{-3}$ \cite{ArkaniHamed, VachaspatiDS}.
In the 2VEV case, the dark sector Higgs and gauge bosons do not directly make up dark matter \cite{ArkaniHamed,ArkaniHamed2}. As a result, there are much less stringent experimental bounds on the model parameters, e.g., if the mixing of the visible sector and the dark sector Higgs particles is small enough, and the dark sector particles are heavy enough, the model is viable.

By setting $q=0\,,C_\mu=0$ we obtain a semilocal model coupled through the Higgs field to a dark scalar field (portal model). 
Assuming that there is an unbroken $\mathbb{Z}_2$ symmetry in the DS, the dark scalar cannot take on a VEV (1VEV case). 
The main interest of such portal models is their minimality in that the dark scalar field itself can be considered as a primary constituent of the dark matter. In the 1VEV case, the gauge boson mass is $m_A^2 = 2$, and the scalar masses are $m_{\phi_1}^2 = 2\beta_1$, $m_\chi^2 = \beta'-\alpha$. Due to the global SU(2) symmetry the field $\phi_2$ stays massless.
Experimental limits on the couplings can be found in Refs.\ \cite{Zee1, Zee2, Beniwal}.
We note that Higgs decays into the dark sector pose rather strong constraints on the coupling $\beta'$ and dark matter density on $m_\chi$.

\section{Visible semilocal strings with a broken symmetry in the dark sector}\label{sec:2VEV}
Straight string solutions in a two-component extended Abelian Higgs model with both fields having a non-zero VEV have been considered for two charged fields in Refs.\ \cite{Erice, BabaevF, BS} and for one charged and one neutral in Refs.\ \cite{FLCC, FLCC2}. In the case of two electrically charged fields, unless the windings of the two scalar fields agree, the energy per unit length of such strings diverges logarithmically\footnote{We assume that the potential vanishes at its minimum. This can be achieved by the subtraction of a constant from the potential $V$ in Eq.\ (\ref{eq:pot}).}, and their flux is fractional.

In the absence of the gauge kinetic mixing term ($\epsilon=0$) the 2VEV vortices of Refs.\ \cite{FLCC, FLCC2} can be embedded in the model given by Eq.\ \eqref{eq:LagPhi}, by setting $\phi_2=0$. For $\epsilon\ne 0$ the angular component of the DS gauge field also becomes non-zero.
The (straight) string solutions we consider are translationally symmetric in the $z$ direction, and rotationally symmetric in $(x,y)$ plane, corresponding to the Ansatz
\begin{equation}
 \label{eq:2vevAns}
 \phi_1 = f(r) \e^{in\vartheta}\,,\quad \chi = g(r)\,,\quad A_\vartheta = na(r)\,,\quad C_\vartheta = c(r)\,,
\end{equation}
where $r,\vartheta$ are polar coordinates in the plane and the other field components ($\phi_2$, $A_r$, $A_z$, $C_r$, $C_z$) vanish.
Using the field equations (\ref{eq:radeq}) one obtains easily that the energy (\ref{eq:edens}) is a monotonously increasing function of the dark charge $q$ [see Eq.\ (\ref{eq:Ergq2})]. The derivative w.r.t.\ the gauge kinetic mixing is given by
\begin{equation}
 \frac{\partial E}{\partial \epsilon} = -2\pi n\int_0^\infty \d r \frac{a' c'}{r}\,,
\end{equation}
which vanishes at $\epsilon=0$, since in that case the field equation for $c(r)$ in Eq.\ (\ref{eq:radeq}) becomes homogeneous and a standard maximum principle argument implies $c(r)\equiv0$. Expanding the fields in a power series of $\epsilon$ [see Eq.\ (\ref{eq:epsexpan})], the energy of the vortex can be written as
\begin{equation}
 \label{eq:Ergeps}
E=E_0 + \epsilon^2 E_2 +O(\epsilon^3)\,,\quad \text{where}\quad E_2 = - 2\pi (2n-1) \int_0^\infty \d r \frac{f_0^2(1-a_0)c_1}{r}\,.
\end{equation}
At $\beta_1=2$, $\beta_2=3$, $\beta'=2$, $\alpha=2.1$, and $q^2=1$
Eq.\ (\ref{eq:Ergeps}) yields an excellent approximation up to $\epsilon \lesssim 0.2$. Moreover $E_0/(2\pi) = 0.906$ and the correction is $E_2/(2\pi)=-0.089$.

A further approximation is to consider the $q^2\to 0$ limit [see Appendix \ref{app:radeq}, esp.\ Eq.\ (\ref{eq:c1})], in which case $c_1 \approx a_0$, simplifying the expression for $E_2$:
\begin{equation}
 \label{eq:Ergepsq}
 E_2 = -\pi(2n-1) \int_0^\infty r \d r \left( \frac{a_0'}{r}\right)^2 + \dots\,.
\end{equation}
Remarkably, $E_2$ in Eq.\ (\ref{eq:Ergepsq}) is proportional to the magnetic energy of the unperturbed vortex. At $\beta_1=2$, $\beta_2=3$, $\beta'=2$, $\alpha=2.1$, and $q^2=0.1$, Eq.\ (\ref{eq:Ergepsq}) yields $E_2/(2\pi)\approx -0.177$. For these parameter values,
Eq.\ (\ref{eq:Ergeps}) gives $E_2/(2\pi) = -0.155$, which compares quite favourably.

Next we summarise the main results of our stability analysis of string (or vortex in the plane) solutions
corresponding to Ansatz \eqref{eq:2vevAns}.
The perturbation equations around the straight string solutions are given in Appendix \ref{app:linear}.
Crucially the fluctuation equations for $\delta\phi_2$ resp.\ $\delta\phi_2^*$ decouple from the other components
(and also from each other). This decoupling is related to $\phi_2\equiv0$ for the background solution and to the coupling structure of the DS.
The (only) known instabilities of the semilocal model (without a DS) have been found in the $\delta\phi_2$ sector. We argue that in the semilocal-DS model the only potential instabilities are expected to appear
in the fluctuation equations for $\delta\phi_2$ (and for $\delta\phi_2^*$), at least for ``not too large'' values of $\beta'$, simplifying considerably the stability analysis. Due to the translation symmetry in the $t$ resp.\ $z$ variables the linearised equations for the corresponding vector-field components $\delta A_0$, $\delta C_0$, resp.\ $\delta A_3$, $\delta C_3$, decouple from each other and from the other components.
Exploiting the symmetries of the background string solution the linearised equations for the components $\Psi_\ell=$($\delta\phi_1$, $\delta\phi_1^*$, $\delta A_i$, $\delta\chi$, $\delta\chi^*$, $\delta C_i$) can be reduced to a coupled system of the form
\begin{equation}
 \label{eq:perteqApp0}
 \mathcal{M}_\ell \Psi_\ell = \Omega^2\Psi_\ell\,,\quad\ell=0,1,\ldots\,,
\end{equation}
constituting a system of 8 second order radial ODE's for a given value of the angular momentum $\ell$.
For more details of the small fluctuation equations
we refer to Appendix \ref{app:linear}, and Refs.\ \cite{baacke, Goodband, FL, FLCC, FLCC2}.

The coupled perturbation system \eqref{eq:perteqApp0} is not expected to give rise to instabilities at least for not too large values of $\beta'$, $\epsilon$.
When $\beta'=\epsilon=0$ the string solution reduces to an Abrikosov-Nielsen-Olesen (ANO) one \cite{Abrikosov, NO} in the visible sector, embedded into the semilocal-DS model with $\phi_2\equiv0$ and $\chi\equiv\eta_2$.
Therefore Eqs.\ \eqref{eq:perteqApp0} decouple into the perturbations of the ANO vortex in the visible, and that of the vacuum in the dark sector. 
In the visible sector the lowest eigenvalues are well known to be positive \cite{Goodband} [e.g., for $\beta_1=2$, the lowest bound state eigenvalue is $\Omega^2=1.76$, the lowest continuum state is at $\Omega^2=\mathop{\rm min}(2\beta_1,2)$], while in the DS positivity is rather obvious as we are perturbing around a true vacuum state [continuum above $\Omega^2=\mathop{\rm min}(2\beta_2\eta_2^2, 2q^2\eta_2^2)$]. 
Simple perturbation theoretic arguments show that for $\beta'\ll1$, $\epsilon\ll1$ the spectrum remains positive.
Therefore in this paper we shall investigate only the decoupled fluctuation equations
for $\delta\phi_2$, which can be written as:
\begin{equation}
 \label{eq:pertdeltaphi22VEV}
  -\frac{1}{r}(r s_{2\ell}')' + U s_{2\ell}= \Omega^2 s_{2\ell}\,,\quad\quad U= \frac{(n a-\ell)^2}{r^2} + \beta_1 (f^2 -1) +\beta' g^2 \,.
\end{equation}
For $\ell=0$, the potential in Eq.\ (\ref{eq:pertdeltaphi22VEV}) has a negative valley close to the origin (the core of the vortex), while for $r\to\infty$ it is given as $(n-\ell)^2/r^2$. The existence of negative eigenvalues depends on the depth of the attractive valley.
The stabilizing effect of the scalar condensate comes from making this attractive potential valley shallower. More quantitatively, for a given value of $\beta_1>1$ by increasing  $\alpha$ (remember $\alpha-\beta'>0$) the negative eigenvalue approaches zero and for some  value $\alpha=\alpha_{\rm s}(\beta_1,\beta_2,\beta')$, it actually reaches zero.
For $\alpha > \alpha_{\rm s}$, DC vortices are then stable. Quite importantly, a large value of the coupling $\alpha$ is also compatible with the experimental bounds on the model, which is quite promising for electroweak-dark strings.

\begin{figure}
 \noindent\hfil\includegraphics[angle=-90,scale=.5]{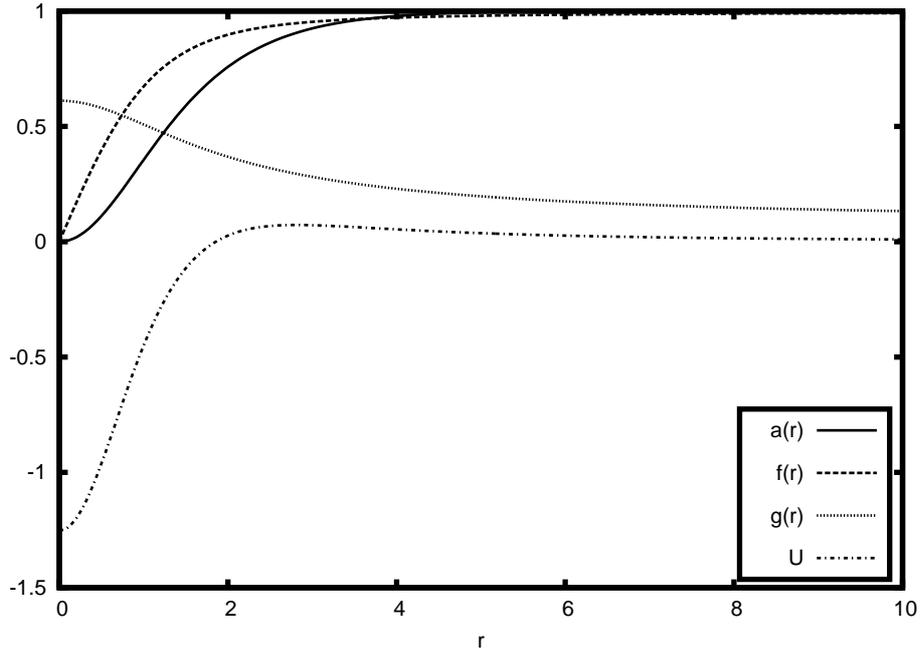}
 \caption{An unstable 2VEV vortex and the potential in its perturbation equation (\ref{eq:pertdeltaphi22VEV}): $\beta_1=2$, $\beta_2=3$, $\beta'=2$, $\alpha=2.011$, and $\epsilon=0$. [For the notation, see Eqs.\ (\ref{eq:2vevAns}) and (\ref{eq:pertdeltaphi22VEV}).]}
 \label{fig:2vev}
\end{figure}

\begin{figure}
 \noindent\hfil\includegraphics[angle=-90,scale=.5]{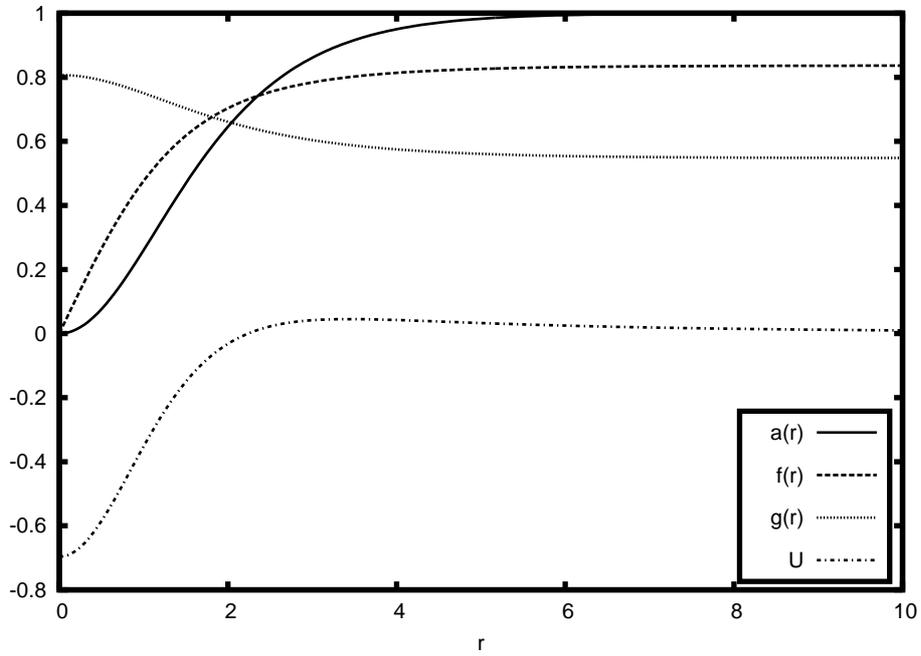}
 \caption{A stable 2VEV vortex and the potential in its perturbation equation (\ref{eq:pertdeltaphi22VEV}): $\beta_1=2$, $\beta_2=3$, $\beta'=2$, $\alpha=2.3$, and $\epsilon=0$. [For the notation, see Eqs.\ (\ref{eq:2vevAns}) and (\ref{eq:pertdeltaphi22VEV}).]}
 \label{fig:2vevs}
\end{figure}

Numerical data are presented in Table \ref{tab:stab2}.
An unstable vortex and the potential in its perturbation equation,  Eq.\ (\ref{eq:pertdeltaphi22VEV}), is shown in Fig.\ \ref{fig:2vev}, and a stable one in
Fig.\ \ref{fig:2vevs}.
As the parameters are tuned, the valley in the potential around the origin becomes shallow, and the bound mode disappears.
Importantly, both the Higgs portal coupling and the gauge kinetic mixing act to stabilise semilocal vortices.

In Table \ref{tab:stab2}, some numerical data of DC vortices are given for $\alpha=\alpha_{\rm s}$, i.e., at the value of $\alpha$ when the change of stability sets in. Note that to larger values of $\epsilon$ correspond lower values of $\alpha_{\rm s}$ (i.e., a larger domain of stability).
One may note that the values of $\alpha_{\rm s}$ decrease of the order of ${\cal O}(10^{-2})$ while $\epsilon$ increases from $0$ to $0.2$. 
Therefore it may appear surprising that the change in the energy is rather small and positive although $\partial E/\partial \epsilon < 0$  while $\epsilon$ changes considerably more than $\alpha$. This effect  can be accounted for by observing that the energy is rather more sensitive to a change in $\alpha$ than to one in $\epsilon$,
e.g.\ at $\beta_1=2$, $\beta_2=5$, $\beta'=2$, $\alpha=4.6$, $E_2 = -0.002 \times 2\pi$ and $\partial E/\partial\alpha \approx -0.344\times 2\pi$.
The relative smallness of $\partial E/\partial \epsilon=2E_2 \epsilon$ as compared to 
$\partial E/\partial\alpha$ can be understood from Eq.\ \eqref{eq:Ergepsq}. Since $a(r)-1$ is exponentially suppressed for large values of $r$
the main contribution to the integral is expected to come from the region of $r<1$, however $f(r)^2={\cal O}(r^2)$ for $r\to0$, accounting for the relative smallness of $E_2$. On the other hand, $\partial E/\partial\alpha={\cal O}(1)$ [see Eq.\ (\ref{eq:Ergalpha}) in Appendix \ref{app:radeq}].

\begin{table}
 \centering
 \begin{tabular}{|c c c| c|c|c|c|c|c|}
  \hline
 $\beta_1$ & $\beta_2$ & $\beta'$ & $\alpha_{\rm s}$ & $E/(2\pi)$   &  $\alpha_{\rm s}$ & $E/(2\pi)$ &  $\alpha_{\rm s}$ & $E/(2\pi)$\\
 \hline
           &           &          & \multicolumn{2}{c|}{$\epsilon=0$} & \multicolumn{2}{c|}{$\epsilon=0.1$} &  \multicolumn{2}{c|}{$\epsilon=0.2$} \\
  \hline\hline
  2    & 5    & 2    & 4.571   & 0.149 & 4.567 & 0.150 & 4.559  & 0.153\\
  2    & 3    & 2    & 2.196   & 0.792 & 2.193 & 0.795 & 2.180  & 0.808\\
  2    & 1.5  & 1.25 & 2.025   & 0.329 & 2.020 & 0.332 & 2.011  & 0.341\\
  \hline
 \end{tabular}
\caption{Stabilisation of 2VEV vortices: the value of $\alpha$ where the vortex becomes stable, and additionally the energy of the vortex at that value of $\alpha$ is displayed. The hidden sector charge is $q^2=1$.}
\label{tab:stab2}
\end{table}

\begin{figure}
 \subfigure[$\beta_2=3,5$]{
 \noindent\hfil\includegraphics[scale=.3,angle=-90]{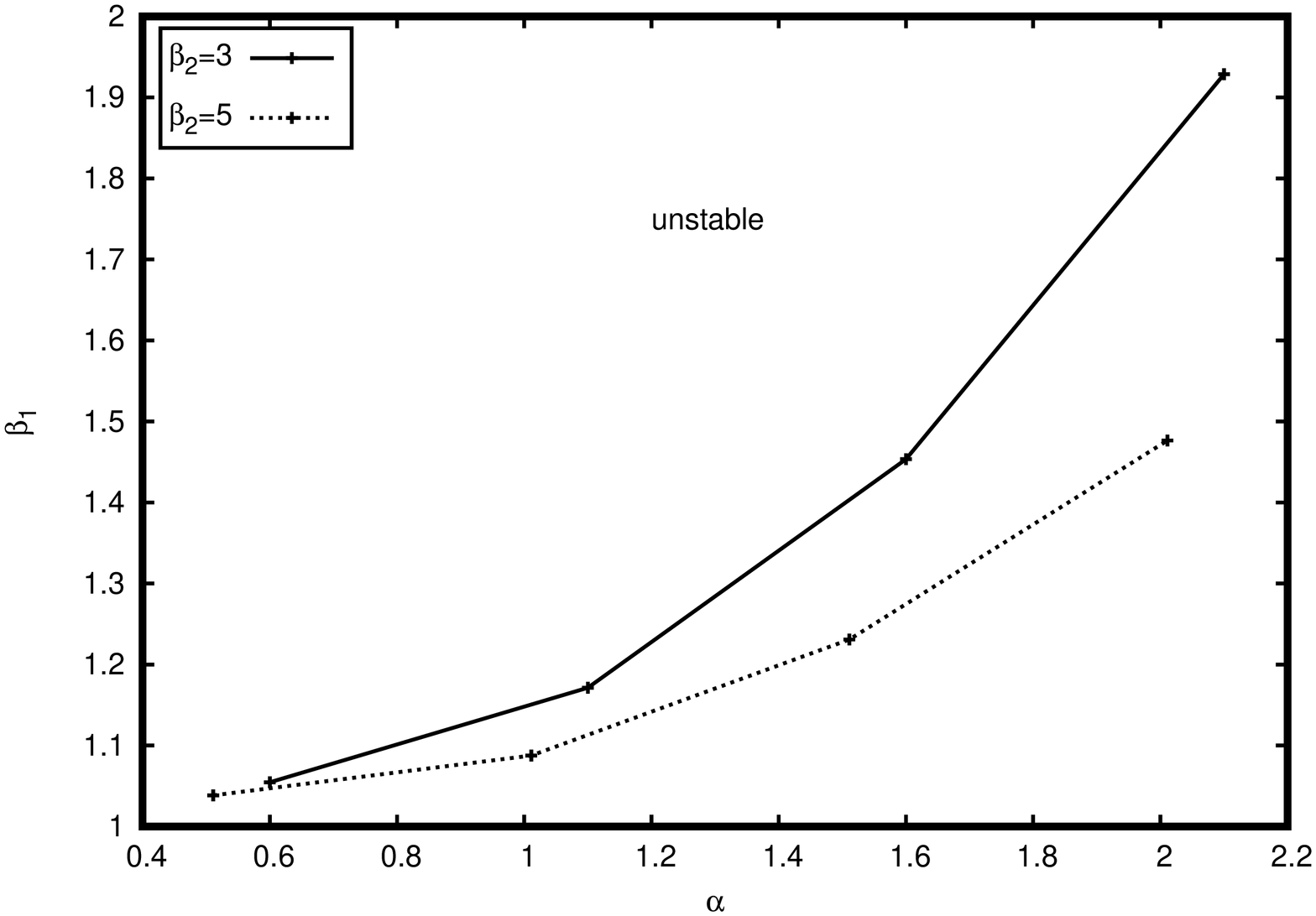}
 }
  \subfigure[$\beta_2=3$]{
 \noindent\hfil\includegraphics[scale=.3,angle=-90]{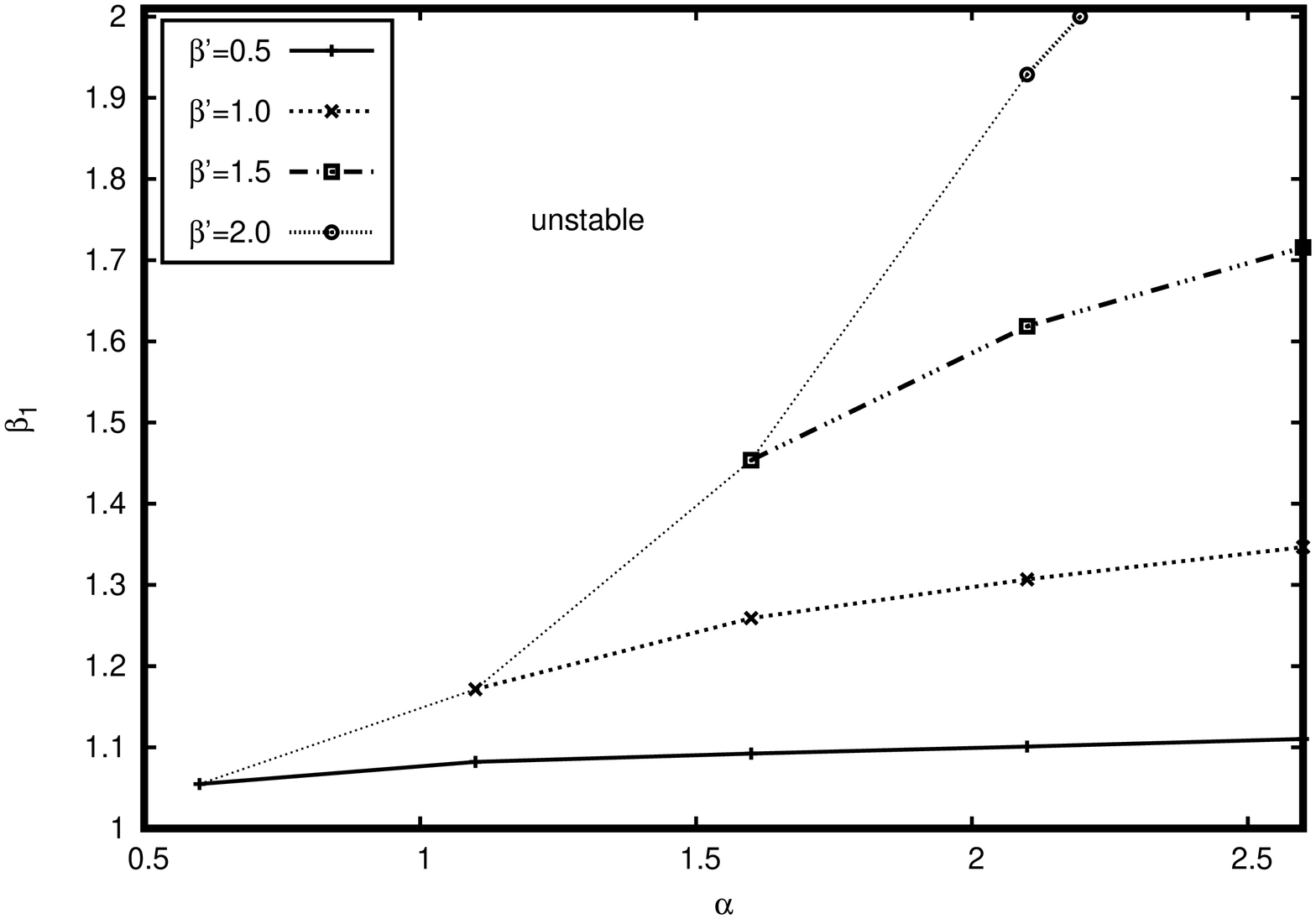}
 }
 \caption{(a) Schematic view of two two-dimensional slices of the domain of stability of DC vortices. (b) Contour plots of the boundary of the domain of stability of DC vortices}
 \label{fig:dstab}
\end{figure}

In Fig.\ \ref{fig:dstab}(a) two-dimensional slices for $\beta_2=3$ and 5 of the domain of stability of DC vortices are depicted schematically. For  values $(\beta_1, \alpha)$ right of the curves, there exist stable DC vortices. Fig.\ \ref{fig:dstab}(a) shows, that the domain of stability increases as $\alpha$ increases, and/or as $\beta_2$ decreases. Fig.\ \ref{fig:dstab}(b) shows additionally the curves separating stable and unstable vortices for fixed values of $\beta'$; these show, that the domain of stability increases as $\beta'$ increases. For better viewing, data points are connected with straight interpolating lines.

\section{Semilocal strings in models with purely scalar dark matter}\label{sec:1VEV}
In Higgs portal models the DS contains only scalar fields, i.e., dark gauge fields are absent. Moreover the VEV of the dark scalars is zero to ensure an unbroken $\mathbb{Z}_2$ symmetry. This case is referred to as 1VEV case in this paper.

The string solutions we consider in this section correspond to the embedding of ``condensate core'' (CC) strings, with $\phi_1=\phi^{({\rm CC})}$, $A_\vartheta = A_\vartheta^{({\rm CC})}$, $\chi=\chi^{({\rm CC})}$ and $\phi_2=0$. CC strings have been studied in Refs.\ \cite{FLCC, FLCC2}. They are the zero current limits of the superconducting strings of Ref.\ \cite{Peter2}.
We refer to these solutions as DC strings.
The linear stability analysis of DC strings is completely analogous to that of the 2VEV case in Sec.\ \ref{sec:2VEV}.
For more details of the perturbation equations, we refer to Appendix \ref{app:linear}.

Again, perturbations of the field $\delta\phi_2$ and $\delta\phi_2^*$ decouple from all other components, and satisfy a Schrödinger-type equation, Eq.\ (\ref{eq:pertdeltaphi22VEV}). The characteristics of the potential $U$ are similar to the one in the 2VEV case; it has a repulsive (centrifugal) contribution determining its $r\to\infty$ asymptotics, and an attractive valley close to the origin, the depth of which depends on the background vortex. 
Therefore the existence of negative eigenvalues depends on the characteristics of the attractive potential valley near the vortex core
in $V$, Eq.\ (\ref{eq:pot}). Since the positive contribution $\beta' g^2$ is of crucial importance, it is illuminating to estimate its
influence. We have found that its parameter dependence is well described qualitatively by approximating the condensate at the vortex core simply by a constant. The minimum of the potential for $\Phi=0$ is at
$|\chi|=\sqrt{\alpha/\beta_2}$. Thus in this crude approximation $\beta' g^2 = \beta' \alpha/\beta_2$ is. Therefore, the larger $\alpha$ and the smaller $\beta_2$ is, the larger the domain of stability of CC vortices is.

\begin{figure}
 \noindent\hfil\includegraphics[angle=-90,scale=.5]{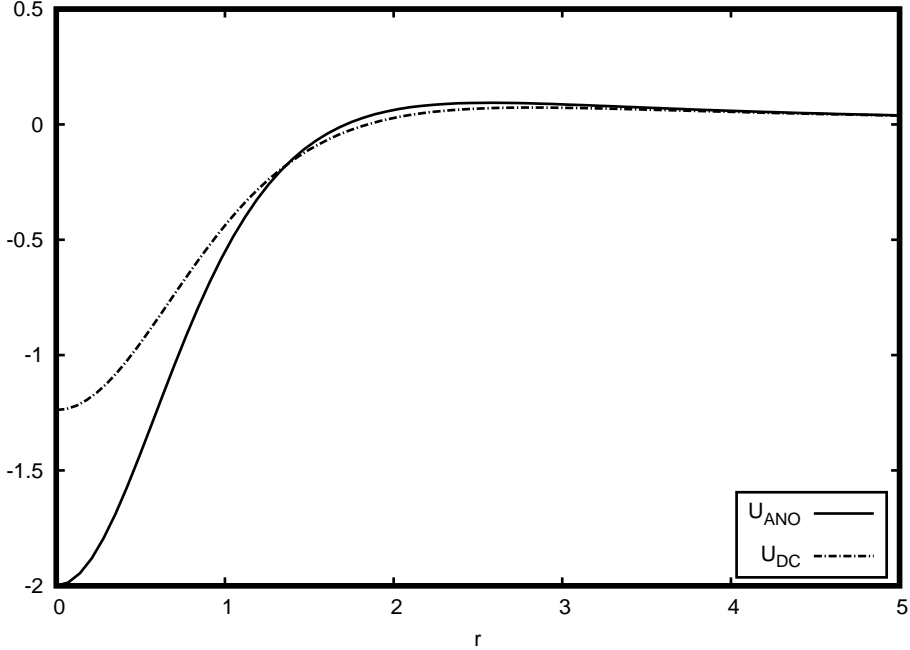}
 \caption{The potentials in the Schrödinger-type equations for perturbations of $\phi_2$ around ANO and DC vortices [Eq.\ (\ref{eq:pertdeltaphi22VEV})], $\beta_{1,2}=2$, $\beta'=2.3$, and $\alpha=2.05$.}
 \label{fig:pot1}
\end{figure}

The reference solution of the semilocal model is the embedded ANO vortex, $\phi_1=\phi^{({\rm ANO})}$, $A_\vartheta=A_\vartheta^{({\rm ANO})}$, $\chi=0$ and $\phi_2=0$.
It is also the unique $z$ independent string for generic $\beta_1$. Embedded ANO vortices have been found to be unstable for $\beta_1 > 1$ \cite{hin1, hin2}. We have found, that DC vortices are stable for $\beta_1 < \beta_{1s}$, where, e.g., for for $\beta_2=6$, $\beta'=2.3$, and $\alpha=2.05$, $\beta_{1s} = 1.247$. Some further numerical data are collected in Table \ref{tab:stab1}.
As it can be inferred from Table \ref{tab:stab1}, the domain of stability of semilocal strings gets significantly extended. A comparison of the potential $U$ for embedded ANO and for DC vortices is shown in Fig.\ \ref{fig:pot1}. The stable DC solutions we found are, however, still  not in the parameter range allowed experimentally (in the SM, $\beta_1 \approx 1.92$), and the Higgs portal coupling here is also too large \cite{Zee1, Zee2, Beniwal}.

The remaining set of fluctuation equations, for the variables $\delta\phi$, $\delta \chi$ and $\delta A_i$, have been investigated numerically in Refs.\ \cite{FLCC, FLCC2}, and no instabilities have been found. This is in contrast with the reference semilocal (embedded ANO) vortex, which does have an instability in this sector. It has been shown, that this does not persist for the CC vortex.

\begin{table}
\centering
\begin{tabular}{|c c c|c|c|}
\hline
$\beta_2$  & $\beta'$ & $\alpha$ & $\beta_{1s}$ & $E/(2\pi)$\\
\hline\hline
3  & 2.3 & 2.05 & 1.615    & 1.0846\\
4  & 2.3 & 2.05 & 1.459    & 1.0630\\
5  & 2.3 & 2.05 & 1.367    & 1.0504\\
6  & 2.3 & 2.05 & 1.247    & 1.0299\\
2  & 2   & 1.85 & 1.805    & 1.1022\\
\hline
\end{tabular}
 \caption{Stabilisation of the strings by the condensate in the 1VEV case. The value of $\beta_1$ and the energy of the vortex at that value of $\beta_1$ is displayed. Embedded ANO strings are stable for $\beta_1 \le 1$. The energy of the ANO vortex for $\beta=2$ is $2 \pi \times 1.1568$, and at $\beta_1=1$, $2\pi$.}
 \label{tab:stab1}
\end{table}

\section{Conclusions}\label{sec:conclusions}
We have investigated the effect of a dark scalar field with Higgs portal coupling and a U(1) gauge field with a gauge-kinetic mixing term on semilocal strings with a local U(1) and
global SU(2) symmetry in the visible sector. The strings considered in this paper have unit winding number with respect to the visible U(1) and zero winding 
number with respect to the dark U(1).
We have found that in a minimal Higgs portal model (with a single dark scalar field), semilocal strings get stabilized 
by a dark scalar condensate at the core of the string. Considering also a dark U(1) gauge field with a gauge-kinetic mixing term 
an additional stabilising effect is found.
These observations open up the possibility of the existence of classically stable dark core electroweak strings.

\subsection*{Acknowledgements}
This work has been supported by the grant OTKA K101709.

\appendix
\section{Scalar masses}\label{app:masses}
To obtain scalar masses in the 2VEV case, we linearise the potential (\ref{eq:pot}) about the vacuum $\phi_1=\eta_1$, $\phi_2=0$ and $\chi=\eta_2$, with
\begin{equation}\label{eq:2VEV}
  \eta_1^2 = \frac{\beta_1\beta_2-\alpha \beta'}{\beta_1\beta_2-(\beta')^2}\,,\quad
  \eta_2^2 = \frac{\beta_1(\alpha- \beta')}{\beta_1\beta_2-(\beta')^2}\,.
\end{equation}
We also introduce new variables, $\delta\phi_1=\phi_1^r + i\phi_1^i$, $\delta\chi=\chi^r + i \chi^i$. The would-be Goldstone bosons, that are later gauged into the longitudinal components of the gauge fields are then $\phi_1^i$ and $\chi^i$. The propagating scalar particles are mixed out of $\phi_1^r$ and $\chi^r$, where their mixing matrix is
\begin{equation}
\label{eq:massmtx}
 M_S = \frac{1}{2}\begin{pmatrix} 4\beta_1 \eta_1^2 & 4\beta' \eta_1 \eta_2 \\ 4 \beta' \eta_1 \eta_2 & 4 \beta_2 \eta_2^2 \end{pmatrix}\,.
\end{equation}
The Higgs particle $H$ and the dark Higgs particle $K$ are related to these as \cite{Vachaspati1}
\begin{equation}
 \label{eq:Higgsmix}
 \begin{pmatrix}
  \phi_1^r \\ \chi^r
 \end{pmatrix}
= \begin{pmatrix}
    \cos\theta & \sin\theta \\
   -\sin\theta & \cos\theta
  \end{pmatrix}
  \begin{pmatrix}
   H \\ K
  \end{pmatrix}\,,
\end{equation}
where
\[
 \tan 2\theta = \frac{2\beta' \eta_1 \eta_2 }{ \beta_2\eta_2^2-\beta_1\eta_1^2 }= \frac{2M_{S12}} { M_{S22} - M_{S11}}\,.
\]
The resulting scalar masses are
\begin{equation}
 \label{eq:scalarmasses}
  M_H^2 = M_{S11} - (M_{S22} - M_{S11})\sin^2 \theta/\cos2\theta\,,\quad M_K^2 = M_{S22} + (M_{S22} - M_{S11})\sin^2 \theta/\cos2\theta.
\end{equation}
The second semilocal component, $\phi_2$ remains massless.

\section{Radial equations}\label{app:radeq}
Inserting the Ansatz (\ref{eq:2vevAns}) into the field equations corresponding to the Lagrangian (\ref{eq:LagPhi}) yields the radial equations
\begin{equation}
 \label{eq:radeq}
 \begin{aligned}
  r \left( \frac{a'}{r} \right)' &= \frac{2}{1-\epsilon^2} f^2 (a-1) + \frac{2\epsilon}{1-\epsilon^2} q^2 g^2 c\,,\\
  r \left( \frac{c'}{r} \right)' &= \frac{2}{1-\epsilon^2}q^2 g^2 c  + \frac{2\epsilon}{1-\epsilon^2} f^2 (a-1)\,, \\
  \frac{1}{r}(rf')'              &= \left[ \frac{n^2 (1-a)^2}{r^2} + \beta_1(f^2-1) + \beta' g^2\right]f\,, \\
  \frac{1}{r}(rg')'              &= \left[ \frac{q^2c^2}{r^2} + \beta_2g^2 -\alpha + \beta' f^2\right]g\,. \\
 \end{aligned}
\end{equation}
The boundary conditions at the origin are demanded by regularity of the fields in the plane, for $r\to 0$, $f \sim f^{(n)} r^n$, $g \to g(0)$,
$a\sim a^{(2)}r^2$ and $c\sim c^{(2)}r^2$.
For $r\to \infty$ we impose $a\to 1$, $c\to 0$, $f\to \eta_1$ and $g\to \eta_2$
in the 2VEV case, and $f\to 1$ and $g\to 0$ in the 1VEV one.

The energy density of a field configuration in the Ansatz (\ref{eq:2vevAns}) is
\begin{equation}
 \label{eq:edens}
 \mathcal{E} = \frac{1}{2}\left[ \left(\frac{n a'}{r}\right)^2 + \left( \frac{c'}{r}\right)^2 - 2 \epsilon n \frac{a'c'}{r^2}\right]
 + (f')^2 + (g')^2 +\frac{n^2 (1-a)^2}{r^2}f^2+\frac{q^2 c^2}{r^2}g^2 + V(f,g)\,,
\end{equation}
where
\begin{equation}
 \label{eq:potrad}
 V(f,g) = \frac{\beta_1}{2}(f^2-1)^2 + \frac{\beta_2}{2}g^4 -\alpha g^2 + \beta' f^2 g^2 -V_0\,,\quad V_0 = -\frac{1}{2}\frac{\beta_1(\alpha-\beta')^2}{\beta_1 \beta_2 - (\beta')^2}\,.
\end{equation}
In Eq.\ (\ref{eq:potrad}), $V_0$ is the term subtracted in the 2VEV case to set the potential to zero at its minimum. In the 1VEV case, no such term is necessary.

As the fields satisfy the Euler-Lagrange equations, in the derivatives of the energy w.r.t.\ the parameters of the model, terms proportional to the implicit derivatives of the fields vanish, and only explicit terms remain, e.g.,
\begin{equation}
 \label{eq:Ergq2}
 \frac{\partial E}{\partial q^2} = 2\pi \int_0^\infty\d r \frac{c^2 g^2}{r} >0\,.
\end{equation}
We spell out explicitly the derivative used in Sec.\ \ref{sec:2VEV} in the 2VEV case,
\begin{equation}
 \label{eq:Ergalpha}
 \frac{\partial E}{\partial \alpha} = -2\pi \int_0^\infty r\d r (g^2-\eta_2^2)\,,
\end{equation}
where the second term is due to the subtraction of $V_0$ in Eq.\ (\ref{eq:potrad}). In the 1VEV case, the derivative is the same as in Eq.\ (\ref{eq:Ergalpha}) without the subtraction of $\eta_2^2$.

A series expansion of the solutions in $\epsilon$ is as follows,
\begin{equation}
 \label{eq:epsexpan}
 a=a_0 + \epsilon^2 a_2 + O(\epsilon^3)\,,\quad f=f_0 + \epsilon^2 f_2 + O(\epsilon^3)\,,\quad g=g_0 + \epsilon^2 g_2 + O(\epsilon^3)\,,\quad
 c=\epsilon c_1 + O(\epsilon^3)\,.
\end{equation}
The resulting equations of motion are obtained for $a_0,f_0, g_0$ by setting $\epsilon=0$ and $c=0$ in the radial equations (\ref{eq:radeq}). The leading order correction, $c_1$ satisfies
\begin{equation}
 \label{eq:c1}
 r \left( \frac{c_1'}{r} \right)' = 2q^2 g_0^2 c_1  + 2 f_0^2 (a_0-1)\,,
\end{equation}
which can be approximated in the limit $q^2\to 0$: In this limit, the right hand side of Eq.\ (\ref{eq:c1}) becomes the same as that of the equation of $a_0$ [see Eq.\ (\ref{eq:radeq})], and therefore $c_1 \approx a_0$.
Although the $q^2\to 0$ limit is not uniform in $r$, the dominant contribution in the energy correction Eq.\ (\ref{eq:Ergepsq}) is expected to come from the core, which is numerically verified.

\section{Linearised equations}\label{app:linear}
For the linearised field equations we use the formalism of Ref.\ \cite{Goodband} (see also Refs.\ \cite{FL, twistedinstab1, twistedinstab2} for applications to multi-component vortices).

For the 2VEV case. in the linearised field equations we introduce a set of new variables for the gauge fields,
\begin{equation}
\label{eq:KmuLmu}
 \delta A_\mu = \frac{\delta K_\mu}{\sqrt{2}\sqrt{1-\epsilon}} + \frac{\delta L_\mu}{\sqrt{2}\sqrt{1+\epsilon}}\,,\quad \delta C_\mu = \frac{\delta K_\mu}{\sqrt{2}\sqrt{1-\epsilon}} - \frac{\delta L_\mu}{\sqrt{2}\sqrt{1+\epsilon}}\,,
\end{equation}
which diagonalise the gauge kinetic terms at the cost of introducing couplings between both gauge fields and both scalars,
\begin{equation}
 \label{eq:ehat}
 {e_-}= \frac{1}{\sqrt{2}\sqrt{1-\epsilon}}\,,\quad{e_+} = \frac{1}{\sqrt{2}\sqrt{1+\epsilon}}\,,\quad{q_-}= \frac{q}{\sqrt{2}\sqrt{1-\epsilon}}\,,\quad{q_+} = \frac{q}{\sqrt{2}\sqrt{1+\epsilon}}\,.
\end{equation}

The linearised equations assume a particularly simple form in the background field gauge \cite{baacke, Goodband},
\begin{equation}
 \label{eq:gauge}
 \begin{aligned}
  F_K &= \partial_\mu \delta K^\mu + i  {e_-}   ( \delta\Phi^\dagger \Phi - \Phi^\dagger \delta\Phi) + i  {q_-}   ( \delta\chi^* \chi - \chi^* \delta\chi)\,,\\
  F_L &= \partial_\mu \delta L^\mu + i {e_+} ( \delta\Phi^\dagger \Phi - \Phi^\dagger \delta\Phi) + i {q_+} ( \delta\chi^* \chi - \chi^* \delta\chi)\,.\\
 \end{aligned}
\end{equation}
The components $\delta K_{0,3}$ and $\delta L_{0,3}$ decouple from the rest of the variables due to the $t,z$ independence of the background, satisfying
\begin{equation}
 \label{eq:K03L03}
 (\square + U_{KK}) \delta K_{0,3} + U_{KL} \delta L_{0,3}=0\,,\quad (\square + U_{LL})\delta L_{0,3} + U_{KL}\delta K_{0,3} =0\,,
\end{equation}
where
\[
U_{KK}= 2{{e_-^2}} \Phi^\dagger \Phi + 2 {{q_-^2}}|\chi|^2\,,\quad U_{KL} = 2{e_-}{e_+} \Phi^\dagger\Phi + 2{q_-}{q_+} |\chi|^2\,,\quad U_{LL} = 2{e_+^2} \Phi^\dagger\Phi + 2 {q_+^2} |\chi|^2\,.
\]

Infinitesimal gauge transformations act on the fields as
\begin{equation}
 \label{eq:infgtrf}
 \delta K_\mu \to \delta K_\mu + \partial_\mu \xi\,,\ \; \delta L_\mu \to \delta L_\mu + \partial_\mu\zeta\,,\ \; \delta\phi_a \to \delta\phi_a + i \phi_a({e_-} \xi + {e_+} \zeta)\,,\ \; \delta \chi \to \delta \chi + i \chi ({q_-} \xi + {q_+} \zeta)\,.
\end{equation}
Due to the residual gauge freedom allowed by the gauge fixing (\ref{eq:gauge}), there are ghost modes, satisfying the equations
\begin{equation}
 \label{eq:ghost}
 (\square + U_{KK}) \xi + U_{KL} \zeta=0\,,\quad (\square + U_{LL})\zeta + U_{KL}\xi =0\,,
\end{equation}
which agree with those of the $0,3$ gauge field components, Eq.\ (\ref{eq:K03L03}), and cancel part of the spectrum, including all modes in the $\delta K_{0,3}$-$\delta L_{0,3}$ sector, therefore, in what follows, we omit these components.

The following Ansatz is compatible with the field equations, due to the cylindrical symmetry of the background,
\begin{equation}
 \label{eq:pertansatz}
 \begin{aligned}
 \delta\phi_1     &=  \e^{i(\Omega t - k z)} \e^{i(n+\ell)\vartheta} s_{1,\ell}(r)         \ \quad+   \e^{-i(\Omega t - k z)} \e^{i(n-\ell)\vartheta}) s_{1,-\ell}(r)\,,\\
 \delta\phi_2     &=  \e^{i(\Omega t - k z)} \e^{i\ell\vartheta}     s_{2,\ell}(r)\quad\ \,\ \quad+   \e^{-i(\Omega t - k z)} \e^{-i\ell\vartheta}     s_{2,-\ell}(r)\,,\\
 \delta\chi\,     &=  \e^{i(\Omega t - k z)} \e^{i\ell\vartheta}    h_{\ell}(r)\,\quad\ \,\,\ \quad+  \e^{-i(\Omega t - k z)} \e^{-i\ell\vartheta}     h_{-\ell}(r)\,,\\
 \delta K_+       &=  \e^{i(\Omega t-k z)}   \e^{i(\ell-1)\vartheta}i t_\ell(r)           \,\ \quad+  \e^{-i(\Omega t - k z)} \e^{-i(\ell+1)\vartheta}it_{-\ell}(r)\,,\\
 \delta K_-       &= -\e^{i(\Omega t-k z)}   \e^{i(\ell+1)\vartheta}i t_{-\ell}^*(r)       \,      -  \e^{-i(\Omega t - k z)} \e^{-i(\ell-1)\vartheta}it_{\ell}^*(r)\,,\\
 \delta L_+       &=  \e^{i(\Omega t-k z)}   \e^{i(\ell-1)\vartheta}i u_\ell(r)           \,\ \quad+  \e^{-i(\Omega t - k z)} \e^{-i(\ell+1)\vartheta}iu_{-\ell}(r)\,,\\
 \delta L_-       &= -\e^{i(\Omega t-k z)}   \e^{i(\ell+1)\vartheta}i u_{-\ell}^*(r)        \,     -  \e^{-i(\Omega t - k z)} \e^{-i(\ell-1)\vartheta}iu_{\ell}^*(r)\,,\\
 \end{aligned}
\end{equation}
where $K_\pm = \exp(\mp i\vartheta)(K_r \mp i K_\vartheta/r)/\sqrt{2}$ and similarly for $L$ (note that $K_\pm^* = K_\mp$).
With the Ansatz (\ref{eq:pertansatz}), the perturbation equation assume the form
\begin{equation}
 \label{eq:perteqApp}
 \mathcal{M}_\ell \Psi_\ell = (\Omega^2 - k^2)\Psi_\ell\,,
\end{equation}
where $\Psi_\ell=(s_{1\ell},s_{1-\ell}^*, s_{2\ell},s_{2-\ell}^*,h_\ell,h_{-\ell}^*,t_\ell,t_{-\ell}^*,u_\ell,u_{-\ell}^*)$. Note that the lowest eigenvalue corresponds to $k=0$, therefore in what follows, we shall only consider such perturbations.
We write the $10\times 10$ matrix operator $\mathcal{M}_\ell$ in Eq.\ (\ref{eq:perteqApp}) as (suppressing all zero entries)
\begin{equation}
 \label{eq:opMell}
 \mathcal{M}_\ell=
   \begin{pmatrix}
D_1           & U_1            &      &           & V             & V'            & {{e_-}} A_1  & {{e_-}} A_1' & {e_+} A_1  & {e_+} A_1'      \\ %s1
U_1           & \bar{D}_1      &      &           & V'            & V             & {{e_-}} A_1' & {{e_-}} A_1  & {e_+} A_1' & {e_+} A_1       \\ %s1*
              &                & D_2  &           &               &               &              &              &            &                 \\ %s2
              &                &      & \bar{D}_2 &               &               &              &              &            &                 \\ %s2%
 V            &  V'            &      &           & D_3           & U_2           & {{q_-}} A_2  & {{q_-}} A_2' & {q_+} A_2  & {q_+} A_2'      \\ %h
 V'           &  V             &      &           & U_2           & \bar{D}_3     & {{q_-}} A_2' & {{q_-}} A_2  & {q_+} A_2' & {q_+} A_2       \\ %h
{{e_-}}  A_1  &  {{e_-}}  A_1' &      &           & {{q_-}}  A_2  & {{q_-}}  A_2' & D_4          &              & U_{KL}     &                 \\ %t
{{e_-}}  A_1' &  {{e_-}}  A_1  &      &           & {{q_-}}  A_2' & {{q_-}}  A_2  &              & \bar{D}_4    &            & U_{KL}          \\ %t*
{e_+} A_1     &  {e_+} A_1'    &      &           & {q_+} A_2     & {q_+} A_2'    & U_{KL}       &              & D_5        &                 \\ %u
{e_+} A_1'    &  {e_+} A_1     &      &           & {q_+} A_2'    & {q_+} A_2     &              & U_{KL}       &            & \bar{D}_5       \\ %u*
%s1                s1*           s2     s2*          h                h*                t           t*               u           u^*
  \end{pmatrix}\,.
 \end{equation}
In Eq.\ (\ref{eq:opMell}), the following notation is used
 \begin{equation}
  \begin{aligned}
    D_1        &=
    - \nabla_r^2
    + \frac{(n(1-  a)+\ell)^2}{r^2}
    + W_1 \,,\\
    D_2        &= %D_s
    - \nabla_r^2
    + \frac{(na-\ell)^2}{r^2}
    + W_2 \,,\\
    D_3        &= -\nabla_r^2 + \frac{\ell^2}{r^2} + W_3\,,\\
    D_4        &= D_K +\frac{(\ell-1)^2}{r^2} \,,\\
    D_5        &= D_L +\frac{(\ell-1)^2}{r^2} \,,\\
\end{aligned}\quad
\begin{aligned}
    \bar{D}_1  &=
    - \nabla_r^2
    + \frac{(n(1 -  a)-\ell)^2}{r^2}
    + W_1 \,,\\
    \bar{D}_2  &=
    - \nabla_r^2
    + \frac{(na+\ell)^2}{r^2}
    + W_2 \,,\\
    \bar{D}_3  &= -\nabla_r^2 + \frac{\ell^2}{r^2} + W_3\,,\\
    \bar{D}_4  &= D_K +\frac{(\ell+1)^2}{r^2} \,,\\
    \bar{D}_5  &= D_L +\frac{(\ell+1)^2}{r^2} \,,\\
\end{aligned}
\end{equation}
with
\[
D_K = -\nabla_r^2 + U_{KK}\,,\quad\quad
D_c = -\nabla_r^2 + U_{LL}\,,
\]
and
\[\begin{aligned}
  W_1 \ \ &= \left(2\beta_1+\frac{1}{1-\epsilon^2}\right) f^2-\beta_1 +\beta'g^2      \,,\\
  W_3 \ \ &=  \left(2\beta_2+\frac{q^2}{1-\epsilon^2}\right) g^2-\alpha + \beta' f^2  \,,\\
  U_2 \ \ &=  \left( \beta_2-\frac{q^2}{1-\epsilon^2}\right) g^2                      \,,\\
  U_{KL}  &= \frac{2}{\sqrt{1-\epsilon^2}} f^2 + \frac{2q^2}{\sqrt{1-\epsilon^2}} g^2  \,,\\
%  U_{LK}  &= U_{KL}                                                                   \,.\\
  A_1 \ \ &= -\sqrt{2}\left(f'-\frac{n f}{r}(1-a)\right)                              \,,\\
  A_1'\ \ &=  \sqrt{2}\left(f'+\frac{n f}{r}(1-a)\right)                              \,,\\
  A_2\, \ &= -\sqrt{2}(g'- qgc/r)                                                     \,,\\
\end{aligned}\hspace{1.5cm}\begin{aligned}
  W_2 \ \ &= \beta_1 (f^2 - 1 ) + \beta' g^2                                          \,,\\
  U_1 \ \ &= \left( \beta_1-\frac{1}{1-\epsilon^2}\right) f^2                         \,,\\
  U_{KK}  &= \frac{2}{1-\epsilon} f^2 + \frac{2q^2}{1-\epsilon} g^2                    \,,\\
  U_{LL}  &= \frac{2}{1+\epsilon} f^2 + \frac{2q^2}{1+\epsilon} g^2                    \,,\\
  V \,\ \ &=  \left( \beta' +\frac{\epsilon q}{1-\epsilon^2}\right) fg                \,,\\
  V'  \ \ &=  \left( \beta' -\frac{\epsilon q}{1-\epsilon^2}\right) fg                \,,\\
  A_2' \  &= \sqrt{2}(g'+ qgc/r)                                                     \,.\\
\end{aligned}
\]

For $\epsilon=0$, the ghost mode equations (\ref{eq:ghost}) decouple. The visible sector one has been solved numerically; it has positive eigenvalues, which change slowly with the parameters. The DS one has a positive potential. Therefore, no modes corresponding to instabilities are cancelled by ghosts.

The formulae presented above also apply for the 1VEV case by setting $\epsilon=q_-=0$, replacing $\delta K_\mu$ with $\delta A_\mu$, and dropping $\delta L_\mu$ and $\zeta$ altogether.

\def\refttl#1{{\sl ``#1''}, }%


\begin{thebibliography}{99}
\newcommand{\NPB}{\sl Nucl.\ Phys.\ \bf B\,}
\newcommand{\PLB}{\sl Phys.\ Lett.\ \bf B\,}
\newcommand{\PL}{\sl Phys.\ Lett.\ \bf}
\newcommand{\PRD}{\sl Phys.\ Rev.\ \bf D\,}
\newcommand{\PRB}{\sl Phys.\ Rev.\ \bf B\,}
\newcommand{\PRL}{\sl Phys.\ Rev.\ Lett.\ \bf}
\newcommand{\PRe}{\sl Phys.\ Rept.\ \bf}
\newcommand{\SJNP}{\sl Sov.\ J.\ Nucl.\ Phys.\ \bf}
\newcommand{\RPP}{\sl Rep.\ Prog.\ Phys.\ \bf}
\newcommand{\CMP}{\sl Commun.\ Math.\ Phys.\ \bf}
\newcommand{\ZPhys}{\sl Zeitschr.\ Phys.\ \bf}
\newcommand{\JHEP}{\sl JHEP\ \bf}


% Kibble's original cosmic string paper
\bibitem{kibbleorig} T.W.B.~Kibble, \refttl{Topology of cosmic domains and strings}
{\sl J.\ Phys.\ A:\ Math.\ Gen.} {\bf 9}, 1387 (1976).

% Cosmic string book & reviews
\bibitem{VS} A.~Vilenkin and E.P.S.~Shellard, {\sl Cosmic strings and other topological defects}, Cambridge University Press, Cambridge, 1994.

\bibitem{kibble} M.B.~Hindmarsh and T.W.B.~Kibble, \refttl{Cosmic strings} {\sl Rep.\ Prog.\ Phys.} {\bf 58}, 477 (1995)
\arxiv{hep-ph/9411342}.

\bibitem{VachaspatiSP} T.~Vachaspati, L.~Pogosian, and D.~Steer, \refttl{Cosmic strings}
{\sl Scholarpedia}, {\bf 10},3168  (2015) \arxiv[atro-ph.CO]{1506.04039}.

\bibitem{Ringeval}
Ch.~Ringeval, \refttl{Cosmic strings and their induced non-Gaussianities in the cosmic microwave background} {\sl Adv.\ Astron.} {\bf 2010}: 380507 (2010).

% semilocal and electroweak review
\bibitem{semilocal} A.~Ach\'ucarro and T.~Vachaspati, \refttl{Semilocal and electroweak strings} {\PRe 327} (2000) 347 \arxiv{hep-ph/9904229}.

%accelerator signature
\bibitem{Nambu} Y.~Nambu, \refttl{String-like configurations in the Weinberg-Salam theory} {\NPB 130}, 505-516 (1977).
\bibitem{Huang} K.~Huang and R.~Tipton, \refttl{Vortex excitations in the Weinberg-Salam theory} {\PRD 23}, 3050 (1981).

% Semilocal cosmic strings
\bibitem{vac-ach} T.~Vachaspati and A.~Ach\'ucarro, \refttl{Semilocal cosmic strings} {\PRD 44} (1991) 3067.
\bibitem{hin1} M.~Hindmarsh, \refttl{Existence and stability of semilocal strings} {\PRL 68} (1992) 1263
\bibitem{hin2} M.~Hindmarsh, \refttl{Semilocal topological defects} {\NPB 392} (1993) 461-492 \arxiv{hep-ph/9206229}.



% electroweak string instability
\bibitem{JPV1} M.~James, L.~Perivolaropoulos, and T.~Vachaspati,
\refttl{Stability of electroweak strings} {\sl Phys.\ Rev.} {\bf D46}, R5232 (1992).

\bibitem{JPV2} M.~James, L.~Perivolaropoulos, and T.~Vachaspati,
\refttl{Detailed stability analysis of electroweak strings} {\sl Nucl.\ Phys.} {\bf B395}, 534-546  (1993) \arxiv{hep-ph/9212301}.

\bibitem{Perkins} W.B.~Perkins, \refttl{$W$ condensation in electroweak strings}
{\sl Phys.\ Rev.} {\bf D47}, R5224  (1993).

\bibitem{GHelectroweak} M.~Goodband and M.~Hindmarsh, \refttl{Instabilities of electroweak strings}
{\sl Phys.\ Lett.} {\bf B363}, 58-64 (1995) \arxiv{hep-ph/9505357}.

% Higgs portal model
\bibitem{SilveiraZee} V.~Silveira, A.~Zee, \refttl{Scalar phantoms} {\sl Phys.\ Lett.} {\bf 161B}, 136-140 (1985).

\bibitem{PW} B.~Patt, F.~Wilczek, \refttl{Higgs-field portal into hidden sectors} \arxiv{hep-ph/0605188} (2006).

% Models of dark matter with a (broken) gauge sector
\bibitem{ArkaniHamed} N.~Arkani-Hamed, D.P.~Finkbeiner, T.R.~Slatyer, and N.~Weiner,
\refttl{A theory of dark matter} {\PRD 79}, 015014 (2009) \arxiv[hep-ph]{0810.0713}.

\bibitem{ArkaniHamed2} N.~Arkani-Hamed and N.~Weiner,
\refttl{LHC signals for a superunified theory of dark matter} {\sl JHEP} {\bf 0812}, 104 (2008) \arxiv[hep-ph]{0810.0714}.

% Dark string physics 1
\bibitem{VachaspatiDS} T.~Vachaspati, \refttl{Dark strings} {\sl Phys.\ Rev.} {\bf D80}, 063502 (2009) \arxiv[hep-ph]{0902.1764}.

% Dark string physics 2
\bibitem{Vachaspati1} J.M.~Hyde, A.J.~Long, and T.~Vachaspati, \refttl{Dark Strings and their Couplings to the Standard Model}
{\PRD 89}, 065031 (2014) \arxiv[hep-ph]{1312.4573}.

% Dark string physics 3
\bibitem{Vachaspati2} A.J.~Long, J.M.~Hyde, and T.~Vachaspati, \refttl{Cosmic Strings in Hidden Sectors: 1. Radiation of Standard Model Particles}
{\sl JCAP} {\bf 09}, 030 (2014) \arxiv[hep-ph]{1405.7679}.

\bibitem{Vachaspati3} A.J.~Long and T.~Vachaspati, \refttl{Cosmic Strings in Hidden Sectors: 2. Cosmological and astrophysical signatures}
{\sl JCAP} {\bf 12}, 040 (2014) \arxiv[hep-ph]{1409.6979}.

\bibitem{Holdom2} C.~Gomez-Sanchez and B.~Holdom, \refttl{Monopoles, strings, and dark matter} {\PRD 83}, 123524 (2011) \arxiv[hep-ph]{1103.1632}.

\bibitem{HindmarshDS} M.~Hindmarsh, R.~Kirk, J.M.~No, and S.M.~West, \refttl{Dark matter with topological defects in the inert doublet model}
{\sl JCAP} {\bf 05} (2015) 048 \arxiv[hep-ph]{1412.4821}.

% Dark string structure
\bibitem{HartmannArbabzadah} B.~Hartmann and F.~Arbabzadah, \refttl{Cosmic strings interacting with dark strings}
{\sl JHEP} {\bf 07} (2009) 068 \arxiv[hep-th]{0904.4591}.

\bibitem{BrihayeHartmann} Y.~Brihaye and B.~Hartmann, \refttl{Effect of dark strings on semilocal strings}
{\sl Phys.\ Rev.} {\bf  D 80}, 123502 (2009) \arxiv[hep-th]{0907.3233}.

\bibitem{BabeanuHartmann} A.~Babeanu and B.~Hartmann, \refttl{Stability of superconducting strings coupled to cosmic strings}
{\PRD 85}, 023518 (2012) \arxiv[hep-th]{1110.5497}.


\bibitem{Fidel1} P.~Arias and F.A.~Schaposnik,
\refttl{Vortex solutions of an Abelian Higgs model with visible and hidden sectors} {\sl JHEP} {\bf 1412}, 011 (2014) \arxiv[hep-th]{1407.2634}.

\bibitem{Fidel2} P.~Arias, E.~Ireson, C.~N\'u{\~n}ez, and F.A.~Schaposnik,
\refttl{${\cal N}=2$ SUSY Abelian Higgs model with hidden sector and BPS equations} {\sl JHEP} {\bf 1502}, 156 (2015) \arxiv[hep-th]{1410.7701}.

\bibitem{Peter2} P.~Peter, \refttl{Low-mass current-carrying cosmic strings}
{\sl Phys.\ Rev.} {\bf D46}, 3322 (1992).


% original model
\bibitem{Witten} E.~Witten, \refttl{Superconducting string} {\NPB 249} (1985) 557--592.

% Two-component CC vortices
\bibitem{Peter} P.~Peter, \refttl{Superconducting cosmic string: Equation of state for spacelike and timelike current in the neutral limit}
{\sl Phys.\ Rev.} {\bf D45}, 1091 (1992).


\bibitem{FLCC} P.~Forgács and Á.~Lukács, \refttl{Vortices with scalar condensates in two-component Ginzburg-Landau systems}
{\sl Phys.\ Lett.} {\bf B762}, 271-275 (2016) \arxiv[hep-th]{1603.03291}.

\bibitem{FLCC2} P.~Forgács and Á.~Lukács, \refttl{Vortices and magnetic bags in Abelian models with extended scalar sectors and some of their applications}
\arxiv[hep-th]{1608.00021}.

% stabilising semilocal strings
\bibitem{VachaspatiWatkins} T.~Vachaspati and R.~Watkins, \refttl{Bound states can stabilize electroweak strings}
{\sl Phys.\ Lett.} {\bf B318}, 163-168 (1993) \arxiv{hep-ph/9211284}.

\bibitem{PerivolaropoulosPlatis} L.~Perivolaropoulos and N.~Platis,
\refttl{Stabilizing the semilocal string with a dilatonic Coupling} {\PRD 88}, 065017 (2013) \arxiv[hep-ph]{1307.3920}.

% Stabilisation with quantum fluctuations
\bibitem{Weigel1} H.~Weigel, M.~Quandt, and N.~Graham, \refttl{Stable charged cosmic strings}
{\PRL 106}, 101601 (2011) \arxiv[hep-th]{1011.2636}.

\bibitem{Weigel2} N.~Graham, M.~Quandt, and H.~Weigel, \refttl{Fermion energies in the background of a cosmic string}
{\PRD 84}, 025017 (2011) \arxiv[hep-th]{1105.1112}.

% Stabilisation with thermal photons
\bibitem{nagabrand} M.~Nagasawa and R.~Brandenberger, \refttl{Stabilization of the electroweak Z string in the early Universe}
{\PRD 67}, 043504 (2003) \arxiv{hep-ph/0207246}.

% gauge kinetic mixing
\bibitem{Holdom} B.~Holdom, \refttl{Two $U(1)$'s and $\epsilon$ charge shifts}
{\sl Phys.\ Lett.} {\bf 166B}, 196-198 (1986).

% Phenomenology of Higgs portal models
\bibitem{Zee1} D.E.~Holz and A.~Zee, \refttl{Collisional dark matter and scalar phantoms}
{\PLB \bf 517}, 239-242 (2001) \arxiv{hep-ph/0105284}.

\bibitem{Zee2} K.~Cheung, Y.-L.S.~Tsai, P.-Y.~Tseng, T.-C.~Yuan, and A.~Zee, \refttl{Global study of the simplest scalar phantom dark matter model}
{\sl JCAP} {\bf 1210}, 042 (2012) \arxiv[hep-ph]{1207.4930}.

\bibitem{Beniwal} A.~Beniwal, F.~Rajec, C.~Savage, P.~Scott, C.~Weniger, M.~White, and A.G.~Williams, \refttl{Combined analysis of effective Higgs portal dark matter models}
{\PRD 93}, 115016 (2016) \arxiv[hep-ph]{1512.06458v2}.


% old 2compt paper
\bibitem{Erice} Á.~Lukács, \refttl{Twisted strings in Extended Abelian Higgs Models} In: A.~Zichichi (ed): {\sl What is known and unexpected at the LHC},
Proceedings of the International School of Subnuclear Physics, Erice-Sicily, Italy, 29 August -- 7 September 2010, World Scientific, 2013.

% Babaev's fractional flux vortices
\bibitem{BabaevF} E.~Babaev, \refttl{Vortices with fractional flux in two-gap superconductors and in extended Faddeev model} {\PRL} {\bf 89}, 067001 (2002)
\arxiv{cond-mat/0111192}.

\bibitem{BS} E.~Babaev and M.~Speight, \refttl{Semi-Meissner state and neither type-I nor type-II superconductivity in multicomponent systems} {\PRB 72} (2005) 180502
\arxiv{cond-mat/0411681}.


% ANO vortex stability
\bibitem{baacke} J.~Baacke and T.~Daiber, \refttl{One-loop corrections to the instanton transition in the two-dimensional Abelian Higgs model}
{\PRD 51}, 795 (1995) \arxiv{hep-th/9408010}.

\bibitem{Goodband} M.~Goodband and M.~Hindmarsh, \refttl{Bound states and instabilities of vortices}
{\sl Phys.\ Rev.} {\bf D52} 4621 (1995) \arxiv{hep-ph/9503457}.

\bibitem{FL} P.~Forgács, Á.~Lukács, \refttl{Instabilities of twisted strings} {\sl JHEP} {\bf 0912} (2009) 064 \arxiv[hep-th]{0908.2621}.

% ANO vortex papers
\bibitem{Abrikosov} A.A.~Abrikosov, \refttl{The magnetic properties of superconducting alloys} {\sl Journal of Physics and Chemistry of Solids}, 2(3), 199-208 (1957).

\bibitem{NO} H.B.~Nielsen and P.~Olesen, \refttl{Vortex-line models for dual strings} {\sl Nucl.\ Phys.}, {\bf B 61} (1973) 45.

% Twisted vortices
\bibitem{FRV1} P.~Forgács, S.~Reuillon, and M.S.~Volkov, \refttl{Superconducting vortices in semilocal models}
{\PRL \bf 96}, 041601 (2006) \arxiv{hep-th/0507246}.

\bibitem{FRV2} P.~Forgács, S.~Reuillon, and M.S.~Volkov, \refttl{Twisted superconducting semilocal strings} {\NPB 751} (2006) 390--418 \arxiv{hep-th/0602175}.


% Twisted vortex instability
\bibitem{twistedinstab1} J.~Garaud and M.S.~Volkov, \refttl{Stability analysis of the twisted superconducting semilocal strings}
{\sl Nucl.\ Phys.} {\bf B799} (2008) 430-455 \arxiv[hep-th]{0712.3589}

\bibitem{twistedinstab2} B.\ Hartmann, P.\ Peter, \refttl{Can type II Semilocal cosmic strings form?} {\sl  Phys.\ Rev.} {\bf D86} (2012) 103516  \arxiv[hep-th]{1204.1270}


%\bibitem{LL3} L.D.~Landau and E.M.~Lifshitz, {\it Course of Theoretical Physics, vol.\ 3, Quantum Mechanics}, Pergamon, Oxford, 1981.



\end{thebibliography}
\end{document}